\documentclass[prd,twocolumn,aps,preprintnumbers,letterpaper,superscriptaddress]{revtex4}
\usepackage{amsmath}
\usepackage{tipa}
\usepackage{amssymb}
\usepackage{graphicx}
\usepackage{bm}
\usepackage{enumerate}
\usepackage{color}
\usepackage[colorlinks = true,
            linkcolor = blue,
            urlcolor  = blue,
            citecolor = blue,
            anchorcolor = blue]{hyperref}
\usepackage{subfigure}

\newcommand{\be}{\begin{equation}}
\newcommand{\ee}{\end{equation}}
\newcommand{\bea}{\begin{eqnarray}}
\newcommand{\eea}{\end{eqnarray}}
\newcommand{\ba}{\begin{array}}
\newcommand{\ea}{\end{array}}
\newcommand{\beal}{\begin{align}}
\newcommand{\eeal}{\end{align}}

\newcommand{\nn}{\nonumber}

\newcommand{\ket}[1]{\left|#1\right\rangle}
\newcommand{\bra}[1]{\left\langle#1\right|}

\def\h0{H^{(0)} }

\def\hri#1#2{\href{http://arxiv.org/abs/#1}{[arXiv:#1]#2}}
\def\hre#1#2{\href{http://arxiv.org/abs/#1/#2}{[arXiv:#1/#2]}}

\def\m{\mu}
\def\n{\nu}
\def\r{\rho}
\def\k{\kappa}
\def\l{\lambda}
\def\s{\sigma}

\begin{document}

\title{Pomeron Interactions from the Einstein-Hilbert Action}
\author{Ioannis Iatrakis}
\email{ioannisiatrakis@gmail.com}
\affiliation{Institute for Theoretical Physics, Utrecht University, Leuvenlaan 4, 3584 CE Utrecht, The Netherlands}
\affiliation{Department of Physics and Astronomy, Stony Brook University,\\ Stony Brook, New York 11794-3800, USA}

\author{Adith Ramamurti}
\email{adith.ramamurti@stonybrook.edu}
\author{Edward Shuryak}
\email{edward.shuryak@stonybrook.edu}
\affiliation{Department of Physics and Astronomy, Stony Brook University,\\ Stony Brook, New York 11794-3800, USA}

\date{\today}

\begin{abstract}
Holographic models of QCD, collectively known as AdS/QCD, have been proven useful in deriving several properties of hadrons. One particular feature well reproduced by such models is the Regge trajectories, both for mesons and glueballs. We focus on scalar and tensor glueballs, and derive an effective theory for the Pomeron by analytic continuation along the leading trajectory from the tensor glueball. It then follows that the Pomeron, as the tensor glueball itself, should possess a two-index polarization tensor, {\em inherited from the graviton}. The three-graviton interaction is deduced from the Einstein-Hilbert action. Using this structure in the cross section of double-Pomeron production of the tensor glueball, we calculate certain angular distributions of production and compare them with those from the CERN WA102 experiment. We find that the agreement is very good for the $f_2(2300)$ tensor glueball candidate. At the same time, other tensor states --  such as $f_2(1270)$ and $f'_2(1520)$ -- have completely different distributions, which we interpret as consequence of the fact that they are not glueballs and thus, in our model, unrelated to the gravitational excitations, which are dual to spin-2 glueballs.
\end{abstract}


\maketitle
\section{Introduction}
\subsection{Pomerons}
Phenomenology of high-energy hadronic collisions dates back to the 1960s, when total, elastic, and diffractive cross sections were first measured and systematically analyzed. Consequences of unitarity and the multi-component nature of intermediate states led to the understanding of the diffractive processes. While first explained from the $s$-channel point of view, the focus of study shifted to the $t$-channel approach soon thereafter. The shape and energy dependence of the amplitudes of scattering processes suggested existence of certain effective objects -- the Reggeons  -- being the analytic continuations of the $t$-channel-exchanged hadrons to the appropriate kinematical domain. Cross-channel unitarities and dualities led to Veneziano amplitudes, which eventually revealed the existence of QCD strings; development of their theory led to the beginnings of string theory.

 The leading term in high-energy behavior of the cross section,
 \be {d\sigma \over dt} \sim s^{\alpha(t)-1}\,, \ee
corresponding to the Regge trajectory with the highest intercept, is known as the Pomeron trajectory. At small negative $t\approx 0$, its linear expansion,
\be  \alpha(t)=\alpha(0)+\alpha' t\,,\ee 
can be used. The original papers of Pomeranchuk assumed a Pomeron intercept of $\alpha(0)= 1$, and that the asymptotic cross section is constant.  Discovery of growing cross sections in 1970s altered the theory into that of a \textit{supercritical Pomeron}, with the intercept slightly larger than one,
\be \alpha(0)=1 +\Delta, \,\,\,\, \Delta\sim 0.08, \ee
where the value of $\Delta$ depends on the scale of the momentum transfer. Universality of the rising part of the total cross sections at large $s$ was tested by $pp$ and $\bar{p}p$ collisions, and is well supported by experimental data.

The phenomenology of high-energy hadronic scattering is significantly complicated by the issue of multi-Pomeron corrections. The proton size is large compared to the natural string scale $(\alpha')^{-1/2}$, masking $s$-dependence of the diffusion process related to Pomeron. Ideally, high-energy collisions of {\em two virtual photons} with large $Q^2$ would be best suited to probe these effects.  The data available from LEP have, unfortunately, too low a $Q^2$; one has to wait for a future high-energy $e^+e^-$ collider. 

There are three theoretical approaches claiming a derivation of the Pomeron amplitude:
\begin{enumerate}[(i)]

\item 
An approach based on the fundamental QCD Lagrangian and {\em perturbative diagrams}. In the leading $\log(s)$ approximation, the scattering is dominated by gluonic ladders.  The Balitsky-Fadin-Kuraev-Lipatov (BFKL) Pomeron \cite{BFKL, Fadin:1975cb, Kuraev:1977fs} explains smallness of  $ \Delta_\text{BFKL}= \mathcal{O}(\alpha_s) $, which approximately  matches the observed ``hard Pomeron" properties at large $t$, although next order corrections to the intercept do not appear to be small. A generic feature of the perturbative approach (with non-running coupling) is that it lacks any dimensional parameters, is conformal, and leads to zero slope in leading order, $\alpha'=0$. Obviously, it cannot be related to Regge trajectories; mesonic, baryonic, or glueball masses; or similar quantities.
  
 \item
 {\em Holographic string-based models}, on the other hand, relate mesonic and baryonic Regge trajectories to rotating open-string states in the bulk spacetime. In this context, the glueballs are related to rotating states of the closed strings. The Reggeon slopes (including the Pomeron's $\alpha'$), in those models, are related to the fundamental parameter string tension in the bulk, which is related to the dual non-perturbative QCD string tension. For a recent derivation and discussions of the Pomeron from stringy holographic perspective, see \cite{arXiv:1202.0831, arXiv:1210.3724,Shuryak:2013sra} .
 
\item 
With the advent of the {\em gauge-string AdS/CFT duality} and models collectively called AdS/QCD, it was naturally questioned whether Reggeons and Pomerons can be effectively described by these constructions. Pioneering papers, such as  \cite{hep-th/0603115}, used such approach to study high-energy scattering processes and related the Pomeron with the ``Reggeized" graviton. Current  AdS/QCD models describe mesonic and glueball states via  quantized (in the holographic dimension) states of few local bulk fields. Conformal invariance of AdS/CFT is broken explicitly, by effective ``confining walls," reproducing dimensional quantities like hadronic masses and the Pomeron slope, $\alpha'$. Regge trajectories, including their ``daughters," naturally appear in such approach. For a recent update, see e.g. \cite{Ballon-Bayona:2015wra}.  
\end{enumerate}

Debates about the role of (i) and (ii) approaches in scattering amplitudes has lead some authors (e.g. Donnachie and Landshoff, \cite{hep-ph/9209205}) to suggest that  the ``hard"  and ``soft"  Pomerons are two different objects, contributing to the scattering amplitude separately, by two additive terms. 

In the holographic models we discuss all hadrons -- and presumably the Pomeron as well -- 
have a single wave function depending on the holographic coordinate $z$,
incorporating the ``hard" (small $z$) and ``soft" (large $z$) parts into a single object. 
This does not prevent existence of different regimes of soft and hard scattering amplitude,
with smooth or non-smooth transition between those. 
 For a recent discussion of the Pomeron profile in various  regimes and its connection to string dynamics and thermodynamics, see \cite{Shuryak:2013sra}.
  
Reggeons and  Pomerons are  complicated non-local objects, and their understanding in terms of basic QCD fields is quite difficult. The non-trivial promise of the holographic approach (iii) is that such complicated objects can perhaps be treated by a dual field theory, operating with a local and weakly coupled set of a few bulk fields. The non-local objects on the boundary, where the gauge theory resides, are obtained after the bulk calculations,
via a direct holographic correspondence. 

In this paper, we will follow this last direction (iii) mentioned above. We will further focus on the question of whether bulk gravity, in the familiar general relativistic form, can be related with the tensor structure of the Pomerons. Our pragmatic philosophy will be to start with the tensor glueball, describe it in certain holographic model, derive the scattering amplitudes in question, and only then  switch to the issue of the Pomeron, treated by an analytic continuation along the leading Regge trajectory, from spin 2 to $1+\Delta$. We thus start with physical, on-shell, tensor glueball $T$ and proceed toward the near-massless $P$, both being certain quantum states of the bulk gravity field.  

While ``sliding" ($t$-dependent)  spin is the basis of the Regge approach, the index structure of the effective vertex can only be formulated with some fixed integer number of indices. 
 In this paper we study a possibility that the Pomeron can be described by symmetric spin-2 tensors.
Our focus will be on the 
Pomeron-Pomeron-Tensor ($PPT$) 6-index vertex.
We  discuss the far-reaching conjecture, namely that {\em those should be described by  the holographic  triple graviton vertices}, following from the Einstein-Hilbert action of a 5 dimensional holographic model of QCD. We will show how one can, in principle, check those, and attempt to do so using known details of double-Pomeron processes observed experimentally.

\subsection{Pomeron Interactions}
Diffractive processes provide an assessment of Pomeron interactions. The Pomeron-Pomeron-Reggeon diagram is related to the so called {\em single diffractive} events, in which the rapidity interval, $\Delta < \log(s)$, is not populated by secondaries. The  {\em double diffractive} events have two unpopulated rapidity intervals,  $\Delta_1+\Delta_2 < \log(s)$. The populated part can be as small as a single hadron at mid-rapidity; we will discuss such events in the second part of the paper.

The  Pomeron-Pomeron-Pomeron ($PPP$) and Pomeron-Pomeron-Reggeon ($PPR$) couplings was extracted from the data in 1970s; for a review of those early works, see \cite{Kaidalov:1979jz}. The $PPP$ coupling value that was extracted,
 \be G_{PPP}(t=0)=0.05\pm 0.01 \,\, \text{GeV}^{-1}\,, \ee
is small on the natural scale of about 1 GeV. Its magnitude is better understood from a  dimensionless combination that enters the Pomeron loops,
\be 
{\left( G_{PPP}\right)^2 \over 4 \alpha'_P }\sim 10^{-2}\,,
\ee
suggesting that the Pomeron loop diagrams can be neglected at current energies (in spite of the fact that this parameter appeared to be enhanced by the factor $s^{\Delta}$, growing with energy). For a recent application of the Pomeron effective field theory and the $PPP$ vertex and diagrams, see e.g. \cite{Gotsman:2013nya}
and references therein. As explained there, another combination of couplings,
\be H=g G_{PPP}  s^\Delta >1\,, \ee
appears in ``fan" (non-loop) diagrams, which do not have a small parameter and thus need to be re-summed. Those diagrams are especially relevant for hadron-nucleus interactions at LHC.

We will not go into details of that, and just emphasize one basic empirical fact:  Gribov's Pomeron effective theory appears to be {\em weakly coupled}, \cite{Gribov:1975kh}. Holographic approaches naturally relate this to the large-$N_c$ suppression of all interactions of the bulk fields. One may wonder what happens when $t$ is large enough, so that pQCD can be used and  the triple-Pomeron vertex can be evaluated explicitly. It is by no means simple to do, but was done; the so-called bare triple vertex \cite{Korchemsky:1997fy} comes from complicated conformal diagrams and provides an answer of the form
\be G_{PPP}\times |t|^{1/2}\sim  \left({g^2 N_c \over 4\pi^2}\right)^2\,, 
\ee 
with rather large coefficient, in no way hinting toward a small $G_{PPP}$.

In the strong coupling regime the issue has been studied using the AdS/CFT correspondence, 
where the bulk 5-dimensional theory is weakly coupled. 
Scattering of vector mesons is modeled by diagrams with vector R-fields, coupled to gravitons.
The closest to our paper is that by Bartels et al \cite{Bartels:2009jb}, in which 6-point 
R-current correlator has been calculated. It includes one diagram with the 3-graviton 
vertex: but unfortunately it was found that it gives no contribution to the kinematic structure
they look for. As we will show below, the double diffractive production cross section
(which is of the second order in this vertex) is not only non-zero, but is even 
successful phenomenologically.

\subsection{The WA102 Experiment and the Tensor Structure of the Pomeron Interactions}
 
Significant progress in Pomeron phenomenology occurred due to CERN WA102 experiment \cite{hep-ex/9803029, hep-ph/0008053, Barberis:2000em}, which studied double diffractive production of $J^P=0^+,0^-,2^+$ hadrons in fixed-target $pp$ collisions at CERN SPS at $\sqrt{s}=29.1$ GeV. This experiment remains, to this day, the main source of information about the double-Pomeron processes. Its analysis has been carried out over the years, resulting in published distributions in both momentum transfers and the angle between them, which we will call $\phi_{34}$, together with the invariant mass distributions for many final-state hadronic channels. Below, we will reproduce some of the most relevant plots from these analyses. 

The collision energy of WA102 experiment is not high enough to discard non-Pomeron contributions as small. Furthermore, since we deal with double diffraction, these contributions can be dominant. Current LHC experiments with elastic/diffractive events and ineleastic collisions are done with different detectors; we call on experimentalists to perform double-diffractive studies of the kind addressed by the WA102 experiment.

The most significant  feature discovered by this experiment was strong dependence on the angle $\phi_{34}$ between the two momenta transferred to the protons. Furthermore, this dependence appears to be qualitatively different for ``mesonic" and ``glueball" hadronic states, both for scalar and tensor states.

The very existence of a nontrivial distribution was a surprise, suggesting that rather radical changes in our views of the effective description of the Pomeron may be needed. Close and  Schuler \cite{Close:1999bi} have famously argued, on the basis of WA102 data, that the Pomeron must have at least a {\em polarization vector}, interacting with some non-conserved current. In particular, they pointed out hat a pseudoscalar vertex can only be made with the 4d antisymmetric $\epsilon$ tensor, and therefore includes vector product of two transverse vectors  $\vec{q}_1 \times \vec{q}_2$. As a result, the $0^-$ production cross section must be proportional to $\sin^2\phi_{34}$. Ellis and Kharzeev \cite{Ellis:1998uy} added an interesting comment: if the Pomeron is described by a vector field, the vertex is of the form 
 $ \epsilon_{\alpha\beta\gamma\delta} G^{\alpha\beta} G^{\gamma\delta}$ , which is the same as in the chiral anomaly and is perhaps related to it. The data from WA102 collaboration confirmed the $\sin^2\phi_{34}$ dependence quite well. It is also important that it is the same for $\eta$ and $\eta'$: the former is a meson while the latter has gluonic admixture via the anomaly. Thus in the pseudoscalar case this angular distribution can \textit{not} be used as a ``glueball filter."

On the other hand, positive $C$-parity of the Pomeron is hard to reconcile with a vector coupling to a current. Therefore, it has been further proposed to use an effective (symmetric) tensor description coupled to the stress tensor, which naturally couples in the same way to a nucleon and an antinucleon. For a relatively recent phenomenological summary and historic references, see \cite{Ewerz:2013kda}.

The holographic approach to Pomeron problem also has its history; let us jump to our direct predecessors which prompted this work.  Anderson, Domokos, Harvey, and Mann had studied Pomeron exchange in $pp$ collisions \cite{Domokos:2009hm} and the pseudoscalar $0^-$ channel production \cite{Anderson:2014jia} in which the Pomeron is modeled by a Reggeizised $2^+$ (graviton) exchange. Since the tensor is symmetric and cannot be convoluted with the antisymmetric $\epsilon$ symbol, the only possibility is that the extra indices  are convoluted directly, from top to bottom of the diagram of Fig. \ref{fig_2to3}, producing another power of $(p_1 p_2)\sim s$, while keeping the $\sin^2\phi_{34}$ distribution intact. These authors also evaluated the absolute value of the cross section, using Sakai-Sugimoto holographic model. Another approach in the study of the holographic diffractive scattering can be found in \cite{Herzog:2008mu} and \cite{Ballon-Bayona:2015wra}.

From the perspective of the holographic models, the most fundamental  hadronic state is the tensor glueball (we will call $T$), since it is described by the gravity field, with its uniquely fixed Einstein-Hilbert action. (Next come scalar glueballs, associated with the bulk dilaton, but those have more model-dependent terms in the action, and may have significant mixing with quark-related meson fields.   Hence, one can additionally study the diffractive production of scalar glueball states, but since this case is considerably more complicated we leave this for a future work.)

Which tensor hadron state is the best approximation to the fundamental tensor glueball? A current consensus is the tensor resonance $f_2(2300)$, as it is called in current particle data tables. There  are several reasons for this conclusion: 
\begin{enumerate}[(i)]
\item Its mass fits well to pure-gauge lattice calculations of glueball spectroscopy.
\item Its width,
\be \Gamma_{2300}=149\pm 40\, \text{MeV}\,, \ee
is rather small for such a high mass. In particular, it is much smaller than that of the somewhat lower tensor state  $f_2(1950)$ of ``normal" magnitude
 \be \Gamma_{1950}=472\pm 18\, \text{MeV}\,. \ee
This small width is taken as a sign of small meson-glueball mixing.
   
\begin{figure}[b!]
\begin{center}
\includegraphics[width=.4\textwidth,angle=0]{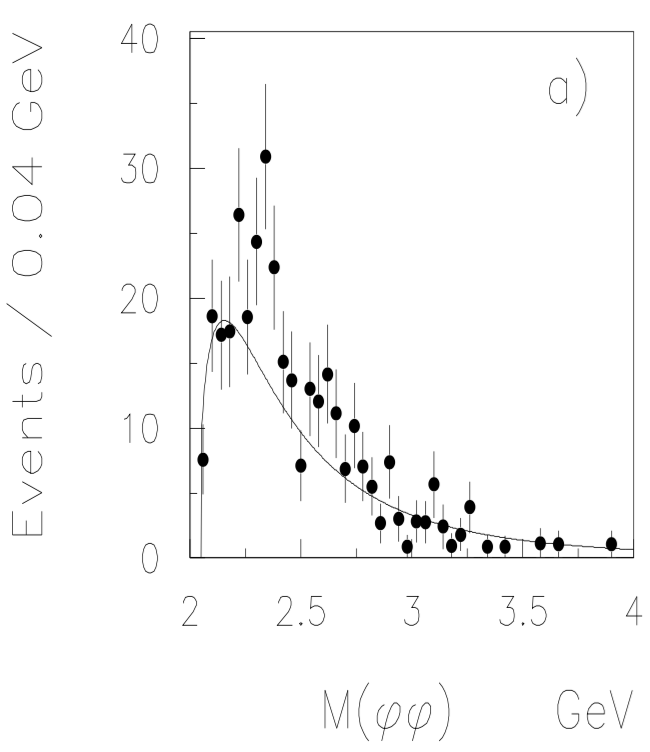}
\includegraphics[width=.4\textwidth,angle=0]{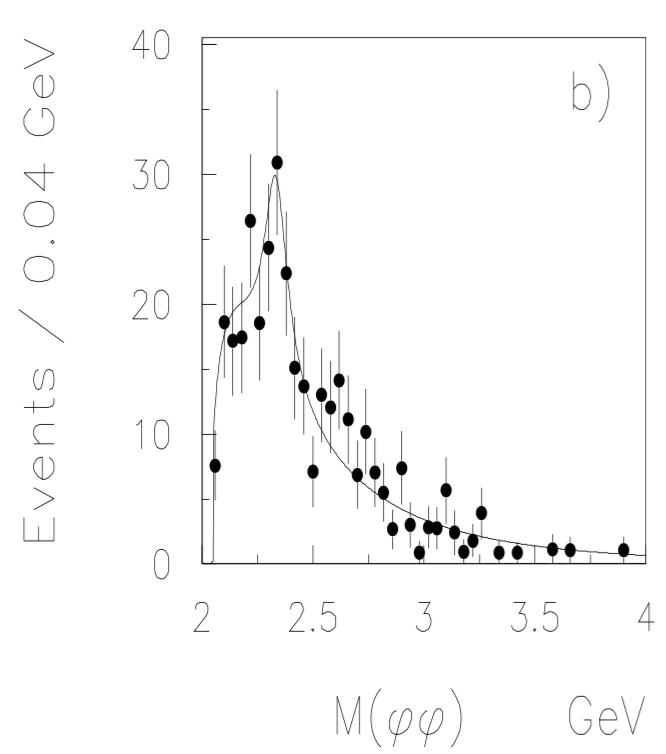}
\caption{ The invariant mass distribution of the $\phi\phi$ channel with total angular momentum $J=2$. The points on both (a) and (b) plots are the same, from the WA102  papers \cite{hep-ex/9803029, hep-ph/0008053}, but are fitted to with different masses of the resonance, 1950 MeV and 2300 MeV, in plots (a) and (b) respectively.}
\label{fig_phiphi}
\end{center}
\end{figure}

\item  This conclusion may appear to be  in contradiction with the WA102 paper, which lists the observed tensor state as $f_2(1950)$ in its title \cite{Barberis:2000em}, but not our preferred state $f_2(2300)$. This is however absolutely \textit{not} the case, as is illustrated by two plots from this paper and reproduced in Fig. \ref{fig_phiphi}. Two plots, (a) and (b) show the same data set, the invariant mass distribution in the $\phi\phi$ spin-2 channel (points) fitted with two masses of the resonance.  It is obvious that the fit at the plot (b) has much better $\chi^2$. We therefore conclude that preference given to the 1950 MeV resonance was perhaps based on some prejudice existing at the time; the data themselves clearly select the other resonance as a clear dominant one, at least in this channel. 
  
 \item Another feature (pointed out in  \cite{Barberis:2000em}) is that the $p_\perp$ and angular $\phi_{34}$ distributions are different for high-mass tensor $T$ as compared to other $J=2$ states  $f_2(1270)$, $f'_2(1520)$ observed in the same experiment. Those states are undoubtedly mesonic (quark-antiquark) states, as a tensor glueball with such a small mass is excluded.
 
 \item Finally,  the situation with $\phi_{34}$ distributions for tensors is similar to that in the scalar channel $J=0$. The distribution of another glueball candidate $f_0(1500)$ is similar to that of $T$, but dissimilar to the scalar mesons. 
 
 \end{enumerate}
 
Compared to the pseudoscalar hadron production channel discussed by \cite{Anderson:2014jia},  the $2^+$ tensor channel is much less restricted by basic symmetries. However, one would argue that precisely because of that, the tensor channel is  more informative. The observation we will make below, that the structure of the Einstein-Hilbert action seem to be in good correspondence with the WA102 data, provides a very non-trivial support to a ``Pomeron-as-graviton" general conjectures.

\section{Kinematics of Double-Diffractive Production}
\label{kinematics}

One starts the introduction of notation with the elastic scattering amplitude: two initial momenta, denoted $p_1,\,p_2$, and the final ones, $p_3,\,p_4$, are used to define the usual Mandelstam invariants $s=(p_1+p_2)^2, t=(p_3-p_1)^2$. Regge kinematics corresponds to $s\gg t$ or near-forward scattering, with Pomeron amplitude already mentioned in the introduction. The maximal cross section is given by an exchange of the highest spin trajectory, that of the Pomeron. The tensor glueball with $J=2$ is the first physical state after the Pomeron, at $t=m_T^2>0$, on this trajectory. 

The actual experiments are mainly done with proton beams, but let us first think of a generic fermion with momentum, $p_\mu$, and spin helicity, $s$, emitting a tensor particle. A proton-proton-glueball vertex is naturally described by the effective action, \be \lambda \int d^4x \, T^{\mu\nu} h_{\mu\nu}\,, \ee with some coupling constant $\lambda$, emitted tensor field $ h_{\mu\nu}$. The stress tensor matrix element between the initial and the final protons is given by,
\be 
\bra{p,s}  T^{\mu\nu} \ket{p',s'}=A(t)\bar u(p',s') {(\gamma^\mu P^\nu+ \gamma^\nu P^\mu) \over 2} u(p,s)\,,
\ee
where sub-leading terms are omitted, so that only one ``large" (symmetrically defined) momentum  $P=(p+p')/2$ is retained. The form factor should, as usual, satisfy $A(0)=1$, and forward matrix element of the stress tensor returns the on-shell nucleon mass.

The upper Pomeron has momentum $k= p_1-p_3 = p_2-p_4$, which is considered small compared to $P$. Furthermore, the two momenta are orthogonal, $k\cdot P=0$, and thus the polarization directions of the glueball are normal to its momenta. Therefore, in the glueball propagator, terms containing $k^\mu,k^\nu$ can all be omitted since they will be multiplied by momentum $P$ from the stress tensor. This allows one to simplify the propagator to
 \be D_{\alpha\beta\gamma\delta} = {\eta_{\alpha\gamma}\eta_{\beta\delta} + \eta_{\alpha\delta}\eta_{\beta\gamma}  
 \over 2(k^2-m_h^2)}\,,
 \ee
 and calculate the elastic cross section to be
 \be {d\sigma \over dt}={\lambda^2 s^4 A^4(t) \over 16\pi (t-m_h^2)^2}\,. \ee
 
The  transferred particle should then be ``Reggeized" according to the Veneziano cross-symmetric form,
\be A\sim {\Gamma(-\alpha(t))\Gamma(-\alpha(u))\Gamma(-\alpha(s)) \over \Gamma(-\alpha(t)-\alpha(s)) \Gamma(-\alpha(t)-\alpha(u)) \Gamma(-\alpha(u)-\alpha(s))}\,, \ee
where the Regge trajectory is assumed to be linear, $\alpha(x)=\alpha(0)+\alpha'(0) x$. By picking only the $t$ pole, one gets the propagator replacement rule,
 \be {1 \over t-m_h^2}\rightarrow {-\alpha' \over 2} {\Gamma(\chi) \Gamma(1-\alpha(t)/2) \over \Gamma(-1+\alpha(t)/2)+\chi)}  e^{-i\pi \alpha(t)/2} \left( {\alpha' s \over 2}\right)^{\alpha(t)-2} \ee 
where important new parameter, $\chi$, in the denominator is defined  by
 \be \chi=\alpha(s)+\alpha(u)+\alpha(t)=4\alpha' m^2+3\alpha(0)\,. \ee
Using such a substitution rule, one gets the standard Pomeron scattering amplitude.

\begin{figure}[htp]
\begin{center}
\includegraphics[width=3.0in,angle=0]{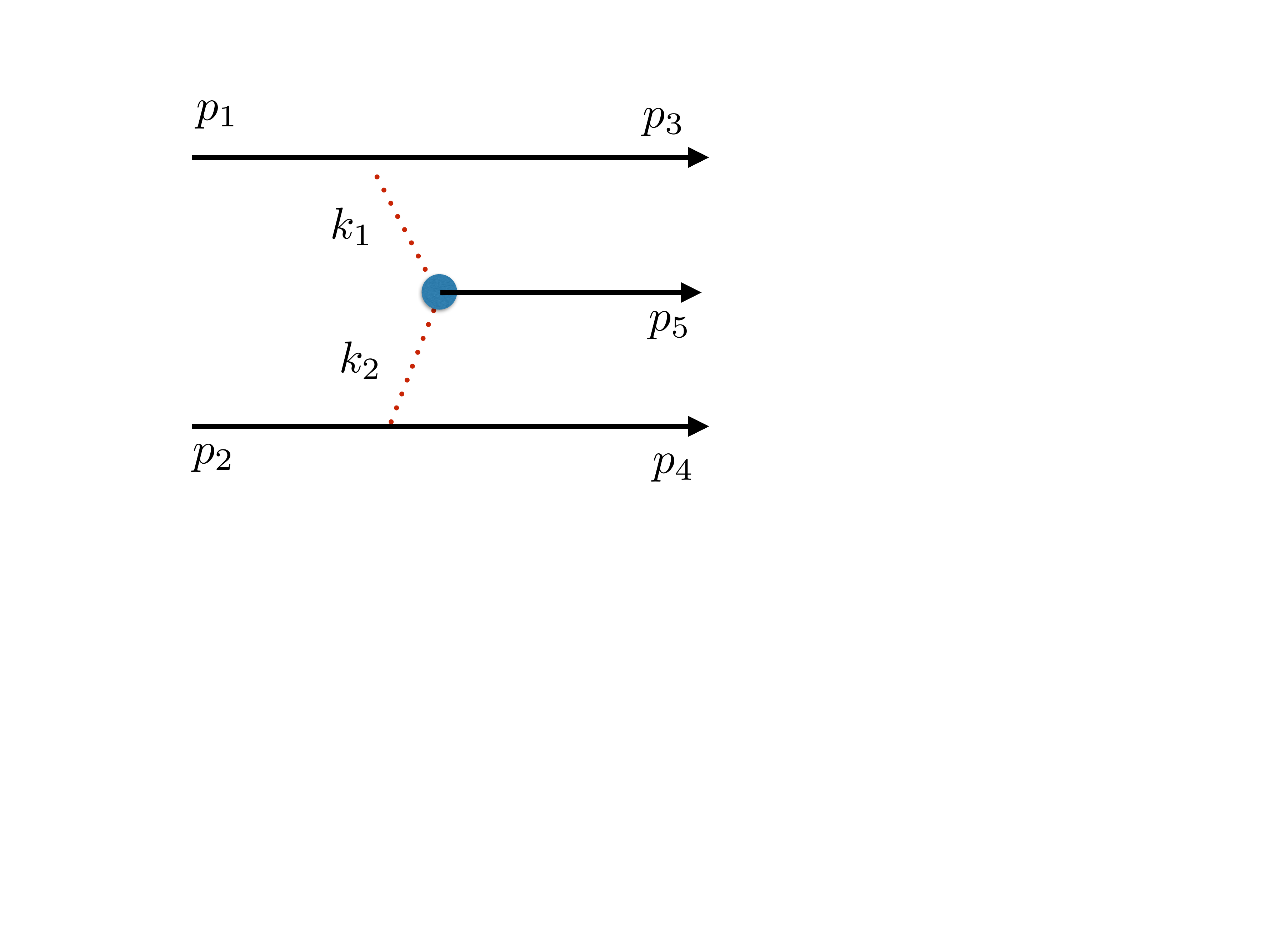}
\caption{The kinematics of the double-Pomeron production.}
\label{fig_2to3}
\end{center}
\end{figure}

The next step is to consider the double-Pomeron production of a single hadron, and then apply the same rules of ``Reggeization" to the lines involved. The kinematics have been discussed in  \cite{Anderson:2014jia} for the case of pseudoscalar hadron; we will follow the same procedure for a tensor glueball. The momenta are defined as shown in Fig. \ref{fig_2to3} (notations coincides with Fig. 1, but not Fig. 2, of \cite{Anderson:2014jia}). Now two momenta transfers, $k_1=p_1-p_3,\,\,k_2=p_2-p_4$, must add up, to form an on-shell hadron with momentum $p_5=k_1+k_2$. We also define 
 \be P^\text{up}=(p_1+p_3)/2,\,\,  P^\text{down}=(p_2+p_4)/2\,, \ee
 which are  orthogonal to the momenta transfer
 \be (P^\text{up} k_1)=(P^\text{down} k_2)=0\,. \ee
 These two vectors, $P^\text{up} \text{ and }P^\text{down}$, written in capital letters, are assumed to be ``large;" their magnitude is $\sim \sqrt{s}$ and are longitudinal (indices 0 and 1, of the beam). Three vectors $k_1,\,k_2,\text{and } p_5$ will be considered of ``medium" magnitude, while the term ``small" will be reserved to the inverse size of the 5$^\text{th}$ holographic dimension. In the case of the LHC, ``large" momenta are $\sim 1000$ GeV, ``medium" are $\sim 1$ GeV and ``small" $\sim .1$ GeV, respectively.
 
Suppose all three secondary particles, $p_3,p_4,\text{ and } p_5$, are detected. In the transverse $yz$ plane these three momenta should sum to zero, so two of them -- that is 4 components -- are not fixed. One global axial rotation is redundant, so there are two transverse momenta, $p_3^{\perp} \text{ and } p_4^{\perp}$, and an angle between them, $\phi_{34}$. Dependence on this angle is determined by the tensor structure of the triple vertex between two Pomerons and a hadron $P(k_1),P(k_2),\text{ and } p_5$, which is defined in an effective  theory of those objects. 
Below we will focus on this dependence, in the model and in the experiment.
  
 \section{The AdS/QCD models, Reggeons and the Pomeron  \label{sec_model}}

In this section, we discuss the accuracy of holographic models in general, which should provide some expectations on whether they should or should not be able to reproduce cross section we want to evaluate. 

Let us start with generic Regge theory and the corresponding trajectories. Semi-classically, the states with large quantum numbers -- large spin $J\gg 1$ and/or large radial quantum number $n\gg 1$ -- correspond to classical rotating or vibrating string states. The ``soft confining-wall models" currently used will all predict linear Regge trajectories, such that $J$ (and/or $n$) $\sim m_{J,n}^2$, the square of the masses of such states. This is well known and we will not discuss it. It is also common knowledge that Regge trajectories for mesons and baryons remain linear, with the same slope, until small $J,\,n,\text{ and }m^2$.

The actual question is whether the Pomeron, with not-too-large spin, $J\approx 1$, and near-zero mass, $t\approx 0$, is located on a linear or a curved Regge trajectory. Since the quantum numbers are not large and the trajectory is neither mesonic nore baryonic one, but a glueball one, one does not a priori know the answer. Glueball trajectories correspond to rotating closed strings, as opposed to open strings for mesons and baryons. The na\"{i}ve picture of non-interacting strings predicts the states to have twice the tension or half the slope, but its accuracy for small $J$ can be questioned.   
 
\begin{figure}[htp]
\begin{center}
\subfigure[]{%
\includegraphics[width=0.4\textwidth]{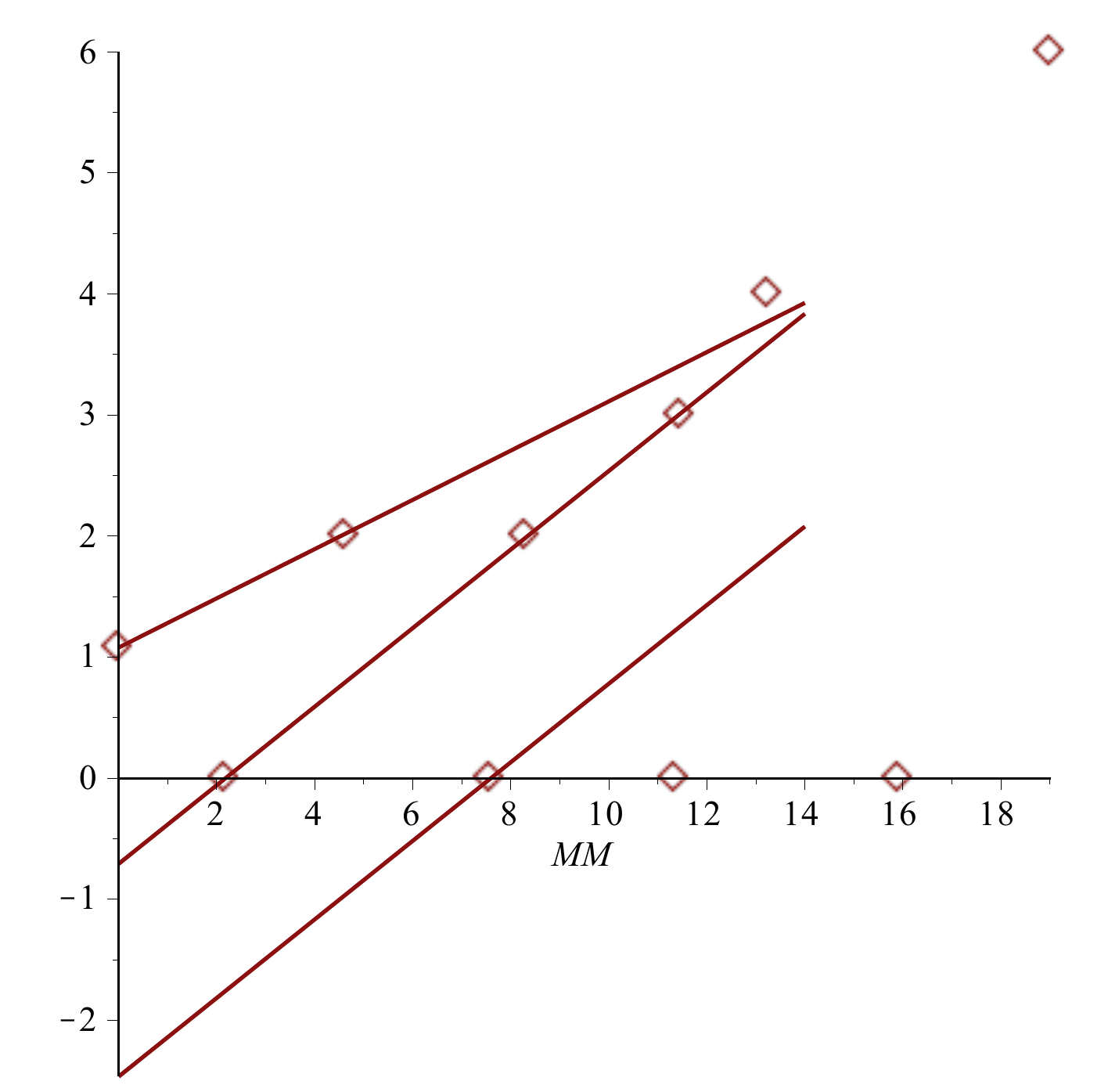}} \quad
\subfigure[]{
\includegraphics[width=0.45\textwidth]{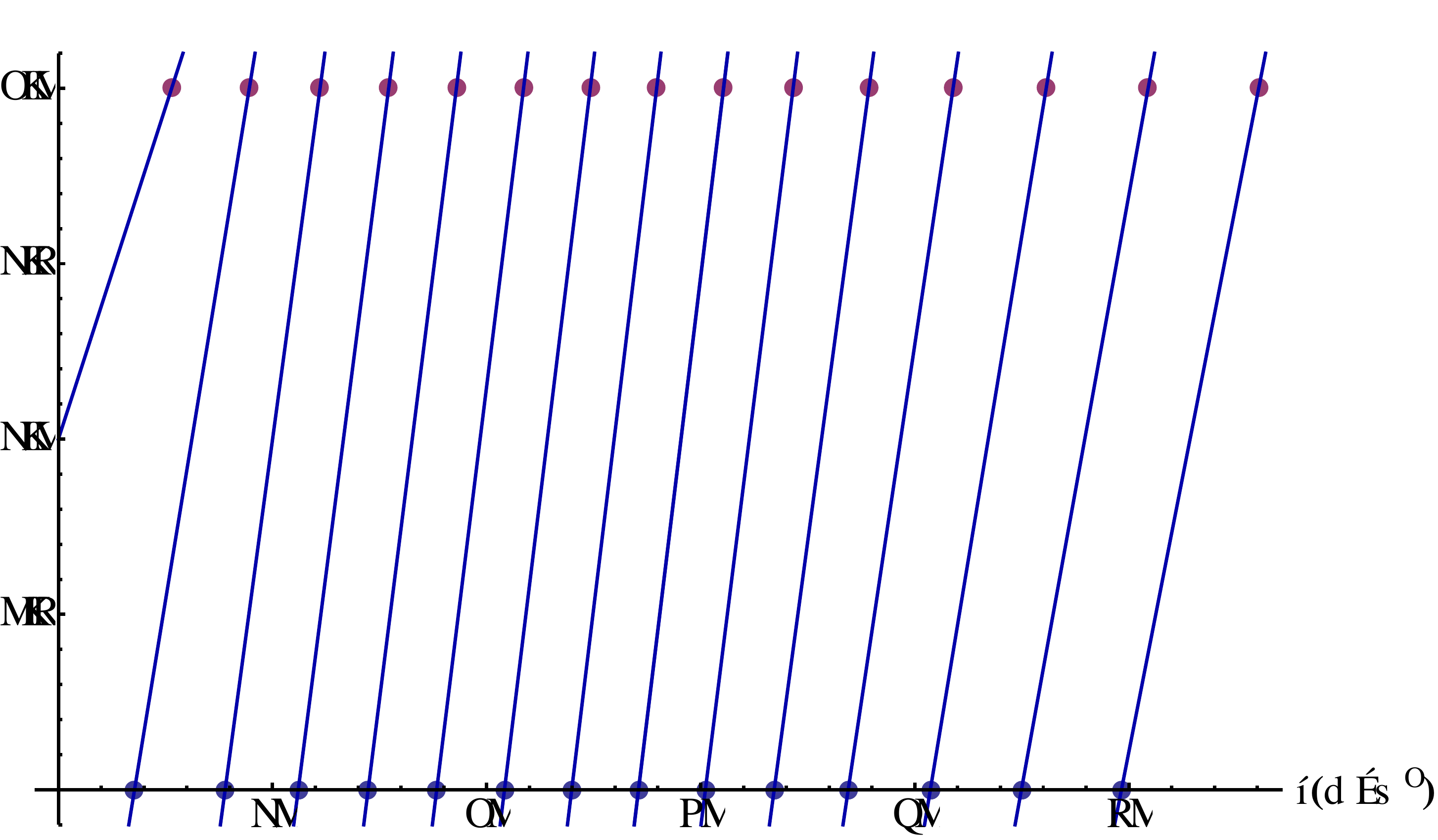}}
\end{center}
\caption{Glueball Regge trajectories, on a plane the total  angular momentum, $J$, versus the mass squared, $MM=t$, in GeV$^2$. The points on the upper plot (a) show the glueball states obtained by numerical lattice simulations \cite{Meyer:2004gx}, while the lower plot (b) shows our own calculation of the lowest scalar and tensor states in the holographic AdS/QCD model.}
\label{fig_Regge}
\end{figure}

Spectroscopy of glueballs is a subject for pure gauge theory, and significant efforts have been made to solve those theories numerically. A compilation of such lattice results were compared to Regge phenomenology \cite{Meyer:2004gx}, and, including the Pomeron, it can be found, for example, in Fig. 5 of \cite{Shuryak:2013sra}, reproduced as \ref{fig_Regge}(a). The points correspond to lattice states of positive parity found in the SU(3) pure gauge theory. While their authors have not discuss or implied that the masses and $J$ of them are related by Regge theory, the reader can see that such trajectory does seem to exist. In particular, the leading one has, apart from the Pomeron, three more states, with $J=2, 4, 6$. A straight line through Pomeron and $2^+$ tensor state $T$ passes close to $4^+$; together with $6^+$ state, one perhaps has a trajectory with some upward curvature. Let us also note that the slope $\alpha'$ of this straight line is also in agreement with that observed in scattering experiments at negative $t$. 
  
The second Regge trajectory through the $J=0,2,3$ states already looks perfectly straight. A few more scalar $J=0$ states were found on the lattice, but their partners with higher spins remain unknown, and thus these two trajectories are all information we currently have.
 
We now ask what are the predictions of the specific holographic models with respect to lowest glueball states and their possible Regge description. Note that there is no need to discuss more recently developed models, such as \cite{Arean:2013tja}, with number of quark flavors $N_f$ as large as the number of colors $N_c$. Since the Pomeron physics is expected to be gluonic, one may take the simpler $N_f=0$ version of this theory.  The lowest glueball states -- scalar and tensor ones -- have already been calculated in this limit. We extended such calculations further, reaching the radial quantum number $n=15$; the results are plotted  in Fig. \ref{fig_Regge}(b). Since we have only pairs of points, we obviously cannot comment on the linearity of these 15 trajectories: but it is seen by eye that a slope has certain variations. The first five scalars and two tensors from the lattice shown in (a) correspond to our calculated masses rather well. The Pomeron location is not calculated, but since we know  empirically that it is close to $J=1$ it is clear from the plot that the slope of the leading -- Pomeron -- trajectory must be quite different and smaller than that of the others. 

 \section{Expanding the Bulk Action to the Cubic Terms  \label{sec_vertex}}

We use, as our effective description of the Pomeron interactions, a model in the class of Improved Holographic QCD (IHQCD) models; see \cite{Gursoy:2007er} for $N_f=0$  and \cite{Jarvinen:2011qe, Arean:2013tja} for $N_f \sim N_c$ . In particular, we use the model for finite $N_f$ in order to compare our results to experimental data of strong interactions, even though we do not expect large contribution from the flavor sector of the model. 
 
The masses of the hadronic states are calculated holographically from the bulk Lagrangians expanded to the second order in perturbations of the bulk fields, on top of the static background minimizing the action.  The main task we perform in this section is expanding the Lagrangian -- in particular, the Hilbert-Einstein gravity Lagrangian $\sqrt{g}R$ -- to the cubic term in gravity perturbation $h$. The gravitational excitation does not couple to any other bulk fields, since there are no other spin-2 bulk fields in the model. Those include (i) the terms with two derivatives, originated from the curvature $R$;  and (ii) terms without derivatives, originated from the volume element $\sqrt{g}$. 
 
If there are no derivatives, the dimensionality comes from curvature of the 5$^\text{th}$ dimension, which is ``small" in our classification. The same happens if the derivatives are along the holographic coordinate, as it was analyzed at the end of section \ref{kinematics}. If the derivatives have Minkowskian indices 0...3, they correspond to momenta $k_1,\,k_2,\text{ and }p_5$, which are ``medium" in our classification, and are therefore the leading order we keep.
 
The action that we expand is
\be 
S_g=M^3 N_c^2 \int d^5x \sqrt{-G} [R-{4 \over 3} G^{MN}\partial_M \Phi \partial_N \Phi +V(\Phi) ] \,,
\ee 
where $M$ is the 5-dimensional Planck mass. The flavor part will not contribute to the fluctuation analysis so we do not mention it here. The solution for the background metric and dilaton is given in \cite{Arean:2013tja}. The choice of the dilaton potential here is also not important for the fluctuation analysis but only for the ``ground state solution," and we have used the Potentials I choice from \cite{Arean:2013tja}. The background metric, describing the zero-temperature state of the field theory, has the following ansatz:
\be
ds^2=e^{2A(z)}(dz^2 + dx_\mu dx^{\mu})\,,
\ee
with $A(z)$ functions depending on the 5$^\text{th}$ coordinate, $z$, only. 

By performing a conformal transformation, $G_{MN}=e^{2 A} g_{MN}$, the action is rewritten as
\begin{align}
S& =M^3 N_c^2 \int d^5x  e^{3 A} \sqrt{-g} \big[ R_g+12 g^{MN} \partial_M A \partial_N A \nn \\
& -{4 \over 3} g^{MN} \partial_M \Phi \partial_N \Phi +e^{2 A}V(\Phi) \big] \,,
\label{act}
\end{align}

We now consider the fluctuation, $g_{MN}=\eta_{MN}+h_{MN}$. We are interested in the spin-2 excitation of the metric; we choose the gauge where
$h_{zz}=h_{z\mu}=0$, $h_\mu^\mu=0$ and $\partial^\mu h_{\mu\nu}=0$. The cubic term of the expansion is
\begin{align}
S& = M^3 N_c^2 \int d^5x  e^{3 A}  \Big[-{1\over 4} h^{MN}h^{KL} \partial_{M}\partial_N h_{KL}  \nn \\ 
&+{1\over 2} h^{MN}h^{KL} \partial_{N}\partial_L h_{MK} + {1\over 2} h^{MN} \partial_P h_{NL} \partial^P h^L_M  \nn \\
& +{1\over 3} h^{MN}h_{NL} h^{L}_{M} e^{2 A}  V(\Phi)  \Big]  \,.
\label{heom}
\end{align}
The above result, in the conformal limit, agrees with the three graviton vertex in ${\mathcal N}=4$, which was first computed in \cite{Arutyunov:1999nw}. The transverse-traceless graviton field is 
\be
h_{MN}(x,z)=\int{d^4 q \over (2\pi)^4}  e^{i q x} \psi(q,z) \Pi_{MN}^{\mu\nu}(q) h_{\mu\nu}^{(0)}(q^2)\, ,
 \ee
where we have introduced the 4-index tensor in $D$ dimensions
 \be 
 \Pi_{\mu\nu\alpha\beta} ={1\over 2}\left( \Pi_{\mu\alpha} \Pi_{\nu\beta} + \Pi_{\mu\beta} \Pi_{\nu\alpha} \right)- {1 \over d-1} \Pi_{\mu\nu} \Pi_{\alpha\beta} 
 \label{proj}
 \ee
 constructed out of the usual transverse two-index tensor of polarizations transverse to momentum $k$,
  $$  \Pi_{\mu\nu}(k)= \eta_{\mu\nu}-{k_\mu k_\nu \over k^2}. $$
While only the particle $p_5$ is the true on-shell spin-2 state,  from our discussion of the kinematics it however follows that this projector can also be applied to two Pomeron lines as well, since those gravitons are also ``transverse" and ``traceless." 
 
 The coupling of the graviton to the energy-momentum operator of the field theory is
 \be
 \int_{\partial {\mathcal M}} h_{\mu\nu} T^{\mu\nu} \, .
 \ee
 According to holography, the three point function of $T_{\mu\nu}$, \cite{Bzowski:2013sza}, reads

 \begin{align}
 T_{\mu\nu\rho\s\k\l}(q_1,q_2,q_3) & = \langle T_{\m\n} (q_1) T_{\k\l} (q_2) T_{\r\s} (q_3)\rangle \nn \\
 & ={\delta^3 S_\text{on-shell} \over h_{\m\n}^{(0)}(q_1) h_{\k\l}^{(0)}(q_2) h_{\r\s}^{(0)}(q_3)}\,.
\end{align}
Taking the graviton to be on-shell, the third term of  Eq. \eqref{heom} vanishes. We also ignore all terms which have no derivatives (such as the fourth term of Eq. \ref{heom}), since they do not contribute in the kinematic limit which we consider, see section (\ref{kinematics}). The resulting expression for the three point 6-index interaction vertex is
\begin{align}
\label{3ptf}
& T_{\mu\nu\rho\s\k\l}(q_1,q_2,q_3)=(2 \pi)^4 \delta^4(q_1+q_2+q_3) M^3 N_c^2 \nn \\
& \int_\epsilon^\infty dz \, e^{3 A(z)} \psi(q_1,z) \psi(q_3,z) \psi(q_3,z) 
 \Big[-{1 \over 2} q_{3 \, b} q_{3 \, d}
\nn \\
&  \big( \Pi_{\m\n}^{ab} (q_1) \Pi_{\r\s}^{cd}(q_2) \Pi_{\k\l \, ac}(q_3)
+ \Pi_{\m\n}^{ab} (q_1) \Pi_{\k\l}^{cd}(q_2) \Pi_{\r\s  \, ac}(q_3)   \nn \\
& +  \Pi_{\k\l}^{ab} (q_1) \Pi_{\m\n}^{cd}(q_2) \Pi_{\r\s   \, ac}(q_3) 
+  \Pi_{\r\s}^{ab} (q_1) \Pi_{\m\n}^{cd}(q_2) \Pi_{\k\l   \, ac}(q_3) \nn \\ 
& +  \Pi_{\r\s}^{ab} (q_1) \Pi_{\k\l}^{cd}(q_2) \Pi_{\m\n   \, ac}(q_3) 
  +  \Pi_{\k\l}^{ab} (q_1) \Pi_{\r\s}^{cd}(q_2) \Pi_{\m\n   \, ac}(q_3)  \big) \nn \\
& +{1\over 2}  q_{3 \, a} q_{3 \, b}  \big( \Pi_{\m\n}^{ab} (q_1) \Pi_{\r\s}^{cd}(q_2) \Pi_{\k\l \, cd}(q_3) \nn \\
&+\Pi_{\r\s}^{ab} (q_1) \Pi_{\m\n}^{cd}(q_2) \Pi_{\k\l \, cd}(q_3) \nn \\
&+\Pi_{\k\l}^{ab} (q_1) \Pi_{\m\n}^{cd}(q_2) \Pi_{\r\s \, cd}(q_3) \big)   \Big]+\mathcal{O}(q^0) \, .
\end{align}
The terms shown above are the ones which contribute to the relevant amplitude for the double Pomeron tensor glueball production, see section (\ref{kinematics}). Even though, the dilaton potential, which is responsible for the non-conformality of the model, does not directly enter the three point function in this limit, it affects the solution for the metric scale factor, $A(z)$, and the glueball wavefunctions, $\psi(q,z)$.
The  equation for the wave function, $\psi$, of the transverse-traceless graviton is found by expanding the action (\ref{act}) to quadratic order \cite{Kiritsis:2006ua, Arean:2013tja}. It reads
\be
\psi(q,z)''+3 A^{\prime}(z) \psi'(q,z)-q^2 \psi(q,z)=0 \,,
\label{graveom}
\ee
where prime denotes the derivative in terms of $z$. The above equation  provides the discrete spectrum of the spin $2^{++}$ glueballs when it is solved by requiring normalizable solutions both in the IR and UV. The wavefunction corresponding to the pomeron is the solution of the above equation in the low $q^2$ limit, where we can solve (\ref{graveom}) perturbatively in $q^2$. Hence, we consider 
\be
\psi(q,z)=\psi_0(z)+q^2 \psi_1(q,z)\, ,
\ee
where $q^2\ll1$. The zeroth-order solution is

\be
\psi_0(z)=c_1 +c_2 \int_{0}^z e^{-3 A(z')} dz' \,.
\label{0sol}
\ee
 Since $A(z)=-z^2$ as $z \to \infty$, the second solution in (\ref{0sol}) is non-normalizable in the IR, so $c_2=0$ for $q^2=0$. Therefore, $c_1=1$ and $\psi_0(z)=1$. The first-order solution in $q^2$ is found by solving the following inhomogeneous equation

\be
e^{-3 A(z)}  \left(e^{3 A(z)} \psi_1' \right)'- \psi_0=0 \,.
\ee
We then have

\begin{align}
\psi_1= \int_0^z e^{-3 A(z')} \int_0^{z'}e^{3 A(z'')}  dz'' \,dz'\, ,
\end{align}
and the total solution for small $q^2$ reads

\be
\psi(q^2,z)=1+q^2 \int_0^z e^{-3 A(z')} \int_0^{z'}e^{3 A(z'')} dz'' \,dz'\,  + {\mathcal O}(q^4) \,.
\ee
The solution satisfies $\psi(q^2,0)=1$ and is normalizable in the infrared region.
 
 \section{The Double-Pomeron Production of Tensor glueballs}

Now we are ready to collect all the ingredients prepared above and calculate the production cross section. The production amplitude depicted in Fig. \ref{fig_2to3} needs to be squared, and summation over the polarizations of the final particles 3, 4, and 5 needs to be performed. The total structure shematically looks as follows, where we have schematically indicated the indices but suppressed momenta:
 $$ h^\text{upper}_{11'}h^\text{lower}_{22'}T^{11'22'33'} \Pi_{33' 44'} T^{55'66'44'} h^\text{upper}_{55'}h^\text{lower}_{66'} \, ,
 $$ 
where $T$ denotes the graviton three point function, Eq. (\ref{3ptf}), which is contracted with the intermediate pomerons mediating the interaction and $\Pi$ is the projector of the on shell glueball state Eq. (\ref{proj}). 
 Here, we have also suppressed the four stress tensors of the protons coupled to the (tensor) field $h$ of the Pomerons, but we remind the reader that the large momenta of the beam produces powers of the largest invariant $s$.   

Using expressions for each block, we used Mathematica to sum all of the indices, which generated an extremely large general expression for the cross section.  We took a series expansion of this expression in the small ratio of the ``medium" scale -- momenta $k_1,k_2,p_5$ and hadronic masses, all $\mathcal{O}(1 \text{ GeV})$ -- to the ``large scale," $p=\sqrt{s}/2$; the lowest nontrivial power of the small parameter is 4. Unfortunately, even this expression is far too large to be put into a paper. We therefore put in specific numbers, corresponding to the kinematics of the WA102 experiment, along with those of the LHC, and plotted the resulting distributions.

 \begin{figure}[h]
\begin{center}
\subfigure[]{%
\includegraphics[width=.49\textwidth]{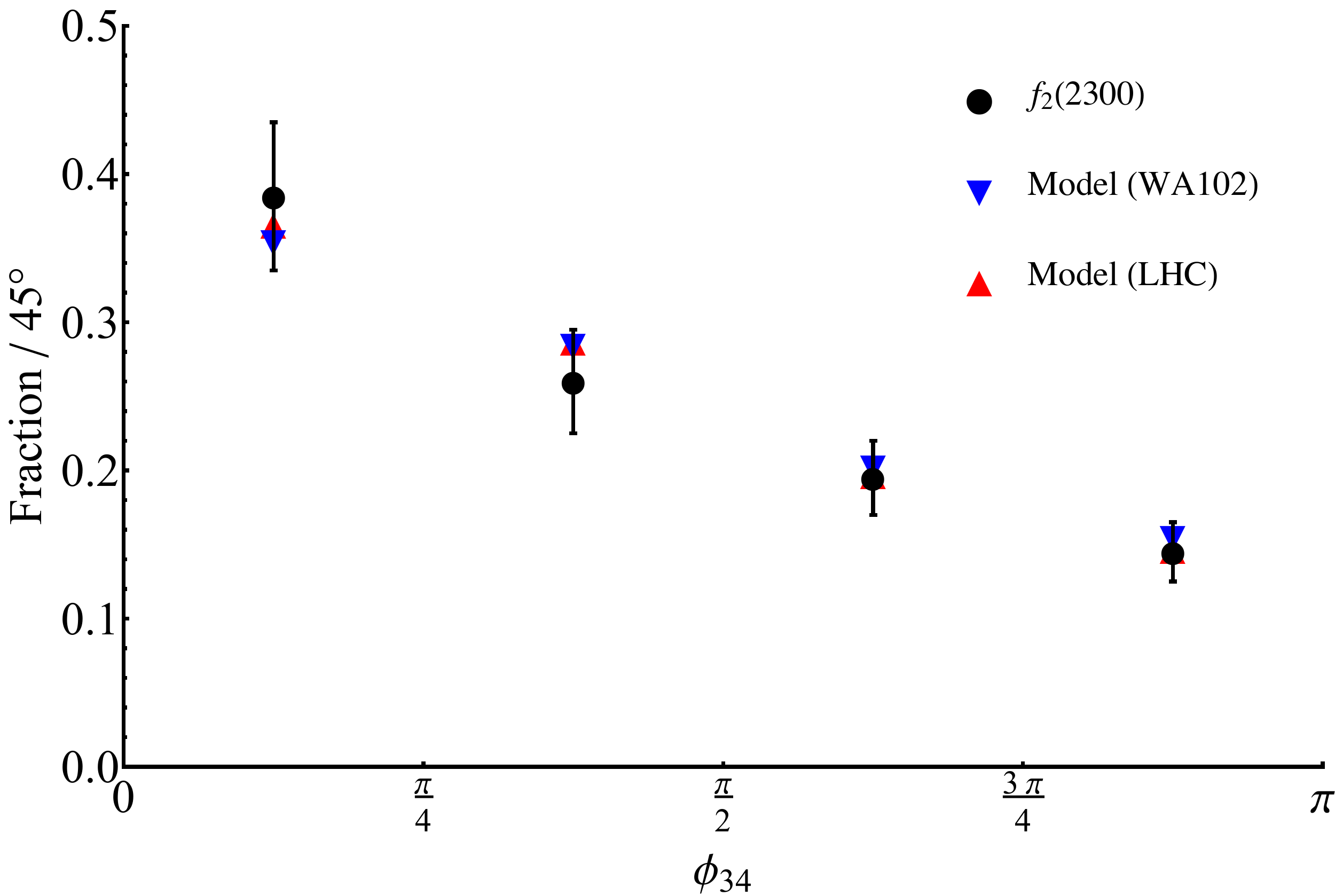}}\quad
\subfigure[]{%
\includegraphics[width=.49\textwidth]{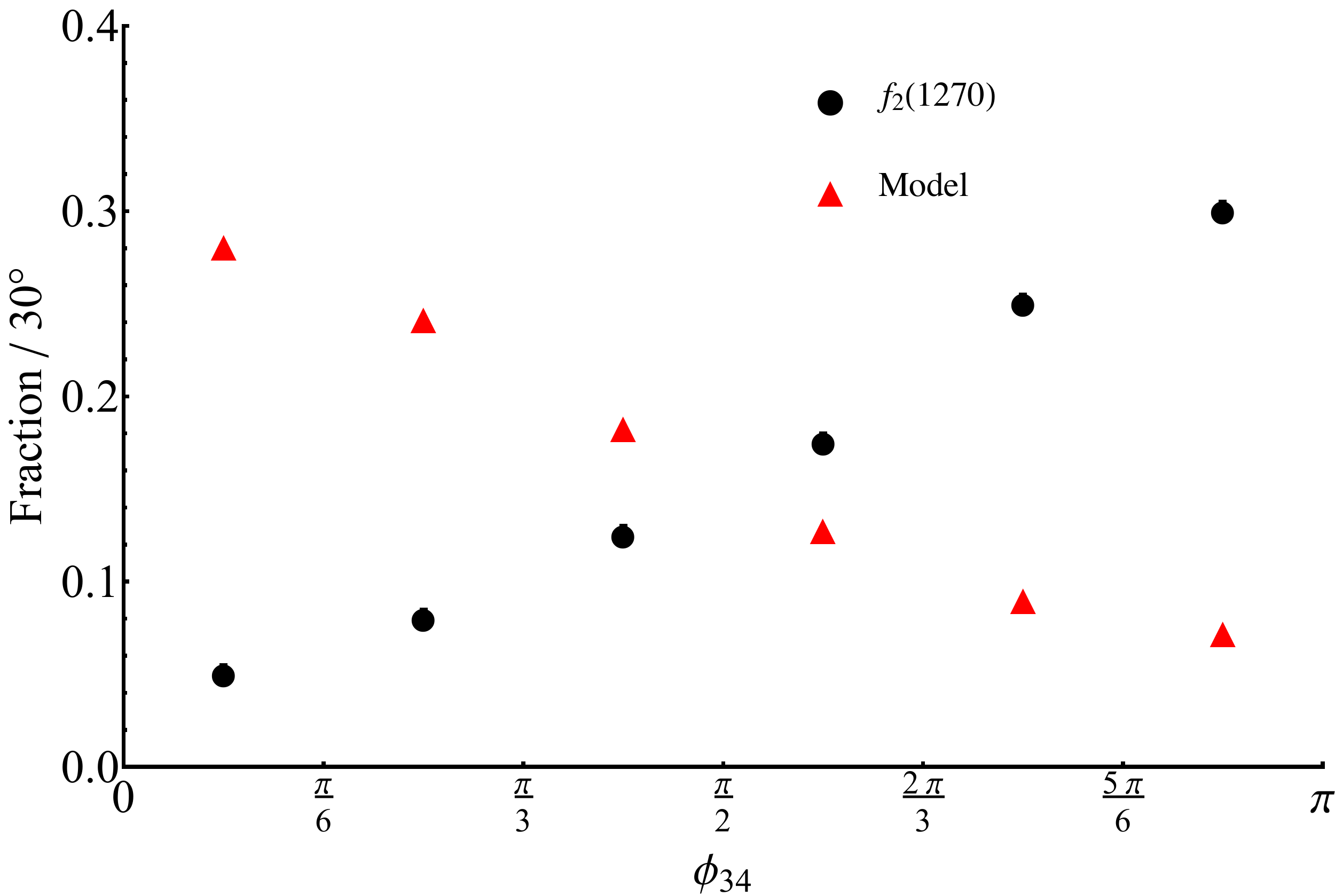}
}
\caption{(a) Double-Pomeron normalized production yield of the 2.3 GeV glueball, $f_2(2300)$ integrated over four azimuthal angular bins. The (black) circles are the data from the WA102 experiment ($\sqrt{s} = 29.1$ GeV) \cite{Barberis:2000em}; downward (blue) triangles are the results of our model at the WA102 energy; and the upward (red) triangles are predicted results at LHC energies ($\sqrt{s} = 13$ TeV). (b)  Normalized production yield of 1.27 GeV spin-2 particle, $f_2(1270)$, integrated over six azimuthal angular bins. (Black) circles are the data of the WA102 experiment \cite{hep-ex/9907055}, and the (red) triangles are our model's predictions.}
\label{PPTfrac}
\end{center}
\end{figure}
  
We note that the transverse momentum distributions for all states are, according to plots in \cite{Barberis:2000em}, simply exponential in momentum transfer squared, $dN/dq^2 \sim \exp(-b q^2 )$. Thus we will use, as representative of the momentum transfer values, their r.m.s. $\bar{q}= 1/\sqrt{b}\approx  0.41$ GeV, where the value of $b$ is from the fit reported in the Table 2 of \cite{Barberis:2000em}. The cross section is, by symmetry, maximal for longitudinal rapidity corresponding to that of the center-of-mass (CM) frame. We thus selected this symmetric kinematics in which the outgoing tensor glueball has zero longitudinal momentum in the CM frame. 
 
The results are seen in Fig. \ref{PPTfrac}. Plotted are the fraction of particles produced at particular separations $\phi_{34}$, integrated over angular bins of width (a) $\pi/4$, and (b) $\pi/6$, and normalized. The angle $\phi_{34}$ is the angle of the transverse momenta $p_3^{\perp}$ and $p_4^{\perp}$ on the transverse plane, see section (\ref{kinematics}). One can see that the results predicted by our model fall firmly within the error bars of the data for the $f_2(2300)$ production collected by the WA102 collaboration. The model's predicted distribution for this particle is only modified slightly as energies are increased to LHC levels. On the other hand, as one can see from Fig. \ref{PPTfrac}(b), the production of another particle, $f_2(1270)$ -- a tensor meson \cite{Klempt:2007cp, Agashe:2014kda}, studied in AdS/QCD in \cite{Katz:2005ir} -- follows a completely different (opposite) trend, with the highest yield occurring in the bin with largest azimuthal separation. We interpret this as a consequence of the fact that this particle is not a glueball, and is, in holographic models, associated with  the bulk quark-related fields coming from the flavor sector of the theory (i.e. the flavor branes) rather than gravity (i.e. the color branes); the triple vertex from the Einstein-Hilbert action, therefore, should not apply. One could add such tensor bulk fields   in the action of the flavor branes in order to describe spin-2 mesons, but this is beyond the scope of this work.
 
 \section{Summary and Discussion}
 
 As discussed in the introduction, the nature of the Pomeron and a quest for its most effective description has occupied the minds of high-energy physicists for half a century. Modern developments in theory have allowed completely new ways of approaching this old problem. Complicated non-local objects -- mesons, glueballs, and perhaps the Pomeron -- are now treated as a holographic images of a relatively simple local field theory in the 5-dimensional bulk.
 
The main idea we followed -- that the Pomeron is a modeled by a symmetric rank 2 tensor -- is, in this approach, quite natural (but by no means new, \cite{Ewerz:2013kda}). Then according to standard AdS/CFT, symmetric spin 2 states that couple to the energy momentum tensor (such as the pomeron) in the boundary field theory are known to be dual to the bulk graviton, \cite{Domokos:2009hm}. In fact, holographic AdS/QCD models had already described the main phases of the matter, their thermodynamics, and gave good effective description of masses and other  properties of the lowest glueballs and mesons, \cite{Arean:2013tja, Gursoy:2007er}. What is, in our opinion, new in this work is an attempt to describe the next order effects: the interactions between those effective objects. 

We have derived the triple-graviton vertex, which follows from the famous Einstein-Hilbert action. The interaction of tensor glueballs was then ``Reggeized," or analytically continued along the Regge trajectory from the on-shell tensor to the Pomeron. We applied the effective description of the Pomeron via tensor field to double-diffractive processes, and modeled effective vertices of the type $PPH$ ($H$ for hadron). The results are compared to experimental data of $f_2(2300)$ production from the WA102 experiment at CERN SPS, with unexpectedly good reproduction of this distribution. 
 
The collision energy of WA102 experiment at CERN SPS was not large enough to avoid a contamination of the non-Pomeron effects, which can, in principle, be significant \cite{Kochelev:2000wm}. Also this experiment was performed  many years ago, in a fixed target setting. One should perhaps seriously consider performing  new generation of double-diffractive experiments at current colliders, RHIC at BNL and LHC at CERN, which have convenient kinematics and large-solid-angle hadronic detectors capable of separating various produced hadrons and their decay channels, to a much better degree than was possible in the WA102 experiment.
\vskip 1cm
{\bf Note added :} After the current work was completed, a new study of the spin structure of the Pomeron \cite{Ewerz:2016onn} appeared. Recent RHIC data  are in good agreement with its tensor structure, and seem to rule out other alternatives.
\vskip 1cm

{\bf Acknowledgements.}
We would like to thank E. Kiritsis for useful discussions. This work was supported in part by the U.S. D.O.E. Office of Science,  under Contract No. DE-FG-88ER40388.
This work is also part of the D-ITP consortium, a program of the Netherlands Organisation for Scientific Research (NWO) that is funded by the Dutch Ministry of Education, Culture and Science (OCW).

\end{document}